\newcommand{\PRE}[1]{{#1}} 
\documentclass[12pt,aps,prd,preprint,tightenlines,
amsmath,amssymb,showpacs,nofootinbib]{revtex4}

\usepackage{enumerate}
\usepackage{color}
\usepackage[pdftex]{graphicx}
\usepackage[colorlinks=true,citecolor=blue,urlcolor=blue]{hyperref}
\graphicspath{{paperfigures/}}	        

\newcommand{\mplanck}{M_{\text{Pl}}}

\newcommand{\gev}{\text{GeV}}
\newcommand{\tev}{\text{TeV}}

\newcommand{\Eqref}[1]{Equation~(\ref{#1})}
\renewcommand{\eqref}[1]{Eq.~(\ref{#1})}
\newcommand{\eqsref}[2]{Eqs.~(\ref{#1}) and (\ref{#2})}
\newcommand{\secref}[1]{Sec.~\ref{sec:#1}}
\newcommand{\secsref}[2]{Secs.~\ref{sec:#1} and \ref{sec:#2}}

\newcommand{\figref}[1]{Fig.~\ref{fig:#1}}
\newcommand{\figsref}[2]{Figs.~\ref{fig:#1} and \ref{fig:#2}}

\begin{document}

\preprint{UCI-TR-2015-16}

\title{\PRE{\vspace*{1.5in}}
MSSM4G: Reviving Bino Dark Matter with \\
Vector-like 4th Generation Particles
\PRE{\vspace*{.5in}}}

\author{Mohammad Abdullah\footnote{maabdull@uci.edu}}
\affiliation{Department of Physics and Astronomy, University of
  California, Irvine, California 92697, USA
\PRE{\vspace*{.4in}}}

\author{Jonathan L.~Feng\footnote{jlf@uci.edu}}
\affiliation{Department of Physics and Astronomy, University of
  California, Irvine, California 92697, USA
\PRE{\vspace*{.4in}}}


\begin{abstract}
\PRE{\vspace*{.2in}} We supplement the minimal supersymmetric standard
model (MSSM) with vector-like copies of standard model particles. Such
4th generation particles can raise the Higgs boson mass to the
observed value without requiring very heavy superpartners, improving
naturalness and the prospects for discovering supersymmetry at the
LHC. Here we show that these new particles are also motivated
cosmologically: in the MSSM, pure Bino dark matter typically
overcloses the Universe, but 4th generation particles open up new
annihilation channels, allowing Binos to have the correct thermal
relic density without resonances or co-annihilation. We show that this
can be done in a sizable region of parameter space while preserving
gauge coupling unification and satisfying constraints from collider,
Higgs, precision electroweak, and flavor physics.
\end{abstract}


\maketitle

\section{Introduction}

Supersymmetric extensions of the standard model are well-motivated by
three promising features that were first identified over three decades
ago.  First, supersymmetry (SUSY) softens the quadratically-divergent
contributions to the Higgs boson mass, reducing the fine-tuning needed
to explain the difference between the electroweak scale and the Plank
scale~\cite{1979Maiani,Veltman:1980mj,Witten:1981nf,Kaul:1981wp}. Second,
the minimal supersymmetric standard model (MSSM) provides the required
new field content to improve the unification of gauge
couplings~\cite{Dimopoulos:1981yj,Sakai:1981gr,Ibanez:1981yh,Einhorn:1981sx}.
And third, with the addition of $R$-parity, supersymmetric extensions
contain stable neutralinos, which are natural candidates for
weakly-interactiong massive particle (WIMP) dark
matter~\cite{Goldberg:1983nd,Ellis:1983ew}.

The lack of direct evidence for supersymmetry, particularly after Run
I of the LHC, has excluded some supersymmetric models, but not
others~\cite{Feng:2013pwa,Craig:2013cxa}, and it remains important to
develop supersymmetric models that continue to have the potential to
realize the original motivating promises.  In this work, we consider
MSSM4G models in which the MSSM is extended to include vector-like
copies of standard model particles. These models have been considered
previously for their promise of raising the Higgs boson mass to the
observed value without extremely heavy superpartners.  We will show
that these models also restore Bino-like neutralinos as excellent dark
matter candidates in a broad range of parameter space that
simultaneously preserves gauge coupling unification and satisfies all
constraints from new physics searches and Higgs, electroweak, and
flavor physics.

In the non-supersymmetric context, the possibility of a 4th generation
of fermions has been considered at least since the 3rd generation was
discovered.  The multiple deaths and rebirths of this idea are nicely
summarized in Ref.~\cite{Lenz:2013iha}. Briefly, in the 1990's a 4th
generation of {\em chiral}, or sequential, fermions was severely
constrained by precision electroweak measurements at LEP, as
parametrized, for example, by the $S$, $T$, and $U$ parameters of
Peskin and Takeuchi~\cite{Peskin:1990zt,Peskin:1991sw}. These
constraints excluded degenerate chiral fermions, which have vanishing
contributions to $T$~\cite{Agashe:2014kda}, but non-degenerate chiral
fermions that contribute to both $S$ and $T$ in a correlated way
remained viable~\cite{Kribs:2007nz}.  The status of chiral 4th
generation fermions changed once again, however, with the advent of
Higgs physics at the LHC.  Since chiral fermions must get their mass
from interactions with the Higgs boson, they contribute to Higgs
production through gluon fusion if they are colored and to Higgs
diphoton decay if they are electrically charged. These contributions
are famously non-decoupling, and current constraints exclude chiral
4th generation fermions up to perturbative values of the Yukawa
couplings.  Although loopholes still exist, for example, in models
with extended Higgs sectors~\cite{BarShalom:2012ms}, even these
possibilities are now severely constrained by the rapid improvements
in precision Higgs measurements, and chiral 4th generation fermions
are now essentially excluded.

The situation is completely different, however, for {\em vector-like}
4th generation fermions.  They can be added in any combination, as
vector-like fermions do not contribute to anomalies, and they may get
masses without coupling to the Higgs boson, so their contributions to
Higgs production and decay do decouple, and they may rather easily
satisfy bounds from precision Higgs measurements. This also means that
they do not contribute to electroweak symmetry breaking effects at
leading order, which keeps them safe from precision electroweak
constraints. Models with vector-like 4th generation fermions therefore
remain viable, and such models have been studied for a variety of
reasons~\cite{Ellis:2014dza}.

In the context of supersymmetry, the possibility of vector-like 4th
generation particles takes on added significance. As is well-known,
the measured Higgs boson mass, $m_h = 125.09 \pm 0.21 \pm
0.11~\gev$~\cite{Aad:2015zhl}, implies there must be large radiative
corrections~\cite{Haber:1990aw,Okada:1990vk,Ellis:1990nz}.  In the
MSSM, this typically requires heavy squarks, which, barring some
explanation, strain naturalness.  But 4th generation fermions and
their scalar superpartners also contribute radiatively to the Higgs
boson mass, reducing the need for very heavy superpartners.  This was
first noted long ago~\cite{Moroi:1991mg,Moroi:1992zk} and has gained
increasing attention through the years as the lower bound on the Higgs
boson mass has grown~\cite{Babu:2004xg,Babu:2008ge,Martin:2009bg,%
  Graham:2009gy,Ham:2010zm,Li:2010hi,Endo:2011mc,Moroi:2011aa,%
  Endo:2011xq,Martin:2012dg,Endo:2012cc,Feng:2013mea,%
  Joglekar:2013zya,Dermisek:2013gta,Fischler:2013tva,Chang:2014aaa,%
  Faroughy:2014oka,Tang:2014bla,Lalak:2015xea,Nickel:2015dna}.  
At the same time, in supersymmetry,
4th generation extensions are highly constrained if one requires that
they preserve gauge coupling unification and raise the Higgs mass
significantly.  These aspects have been discussed at length, for
example, in Ref.~\cite{Martin:2009bg}, where the different
possibilities for vector-like fermions were explored exhaustively with
respect to their ability to increase the Higgs mass, while maintaining
gauge coupling unification and avoiding bounds from electroweak
precision data.

In this study, we show that, in supersymmetry, vector-like 4th
generation particles are also motivated cosmologically.  In many
well-motivated supersymmetric models, renormalization group evolution
or other effects imply that the Bino is the lightest gaugino, and so
it is the lightest neutralino in ``half'' of parameter space (with the
Higgsino being the lightest in the rest of parameter space).  Pure
Binos do not annihilate to $W$ or $Z$ bosons, and they annihilate to
standard model fermions only through $t$-channel sfermions.  For these
annihilation channels to be sufficiently efficient that Binos do not
overclose the Universe, Binos must be lighter than about
$300~\gev$~\cite{Olive:1989jg,Griest:1989zh}. Such light Binos are now
excluded in many cases by results from the LHC.  For example, searches
for gluino pair production, followed by decays to neutralinos, exclude
neutralino masses below 300 GeV, provided the gluinos are lighter than
1.4 TeV and not highly degenerate with the
neutralinos~\cite{Aad:2014wea,Khachatryan:2015vra}.  Light neutralinos
produced in squark decays are similarly
excluded~\cite{Aad:2014wea,Khachatryan:2015vra}.  These bounds have
loopholes. For example, if neutralinos are degenerate with staus to
within 5\%, they co-annihilate in the early Universe and may be as
heavy as 600 GeV without overclosing the
Universe~\cite{Griest:1990kh,Cyburt:2009pg}. Such possibilities are
currently viable, and will be probed completely in the upcoming LHC
run~\cite{Konishi:2013gda,Desai:2014uha,Feng:2015wqa}.  However,
barring such degeneracies and other accidental mass arrangements, Bino
dark matter in the MSSM is now significantly constrained.

Here we will show that vector-like copies of 4th (and 5th) generation
fermions open up new annihilation channels for the Bino, reducing its
thermal relic density to the measured value or below. These new
channels are extremely efficient, with even a single 4th generation
lepton channel dominating over all MSSM channels combined.  Binos are
therefore restored as excellent dark matter candidates in regions of
parameter space where naturalness is improved, gauge coupling
unification is preserved, and all constraints are satisfied. Dark
matter in 4th generation supersymmetry models has been discussed
previously.  In Refs.~\cite{Joglekar:2012vc,Fairbairn:2013xaa}, for
instance, 4th generation neutrinos were considered as dark matter
candidates.  In Refs.~\cite{Moroi:2011aa,Endo:2011xq}, neutralinos
were shown to be viable dark matter candidates when highly degenerate
with co-annihilating sleptons.  To our knowledge (and surprise), there
are no discussions in the literature of the effects of vector-like 4th
generation particles on the thermal relic density of Binos in the
generic, non-co-annihilating case, which is the focus of this study.

The paper is organized as follows. In \secref{model} we present the
particle content, simplifying assumptions, and existing bounds for the
4th generation models we will study.  Simply requiring that the
vector-like 4th generation particles preserve gauge coupling
unification and contribute significantly to the Higgs boson mass
reduces the number of models to consider to essentially two.  We then
examine these two models in detail in \secsref{relic}{higgs}, where we
present out results for the relic density and Higgs mass,
respectively. In \secref{end} we summarize our findings and comment on
the experimental prospects for discovering supersymmetry in these
cosmologically-motivated models.

\section{The Model}
\label{sec:model}

\subsection{Particle Content}

The standard model, supplemented by right-handed neutrinos, includes
quark isodoublets (doublets under the weak isospin SU(2) gauge group)
$Q$, up-type quark isosinglets $U$, down-type quark isosinglets $D$,
lepton isodoublets $L$, charged lepton isosinglets $E$, and neutrino
isosinglets $N$.  Beginning with the MSSM, we add vector-like copies
of these fermions (and their superpartners).  By this we mean adding
both left- and right-handed versions of fermions whose
SU(2)$\times$U(1)$_Y$ charges are identical to one of the standard
model fermions.  As we are only considering vector-like extensions
here, as a shorthand, we will list only one of the chiral fields, with
the chiral partner implicitly included.  Thus, for example, a model
with an extra $Q$ (or {\bf 5}) multiplet implicitly also includes its
chiral partner $\bar Q$ (or ${\bf \bar{5}}$).

Gauge anomalies cancel within each vector-like pair, so there is no
need to add a full generation at once. This would seem to lead to a
Pandora's box of possibilities.  However, the number of models to
consider may be greatly reduced simply by requiring that the new
particles preserve gauge coupling unification and contribute
significantly to raising the Higgs boson mass.

To preserve gauge coupling unification, we begin by considering only
full SU(5) multiplets, that is, {\bf 1}, {\bf 5}, and {\bf 10}
multiplets. Using 1-loop renormalization group equations (RGEs), the
gauge couplings remain perturbative up to the GUT scale with a full
vector-like generation of {\bf 5}+{\bf 10}, but this is not true when
3-loop RGEs are used~\cite{Martin:2009bg}. Thus, gauge coupling
unification reduces the remaining possibilities to either one {\bf 10}
multiplet or one, two, or three {\bf 5} multiplets (plus any number of
singlets).

The {\bf 5} multiplets contain $D$ and $E$ fields. To raise the Higgs
boson mass, these fields must couple to the Higgs field.  The $D$
field would require a $Q$ field, which would bring in an entire {\bf
  10}, ruining gauge coupling unification.  The $E$ field requires
only an $N$, which is consistent with gauge coupling unification.
However, as shown in Ref.~\cite{Martin:2009bg}, perturbativity up to
the GUT scale requires that lepton Yukawa couplings be at most $h =
0.75$.  The contribution of $N_g = 3$ extra generations of
leptons/sleptons to the Higgs boson mass scales as $N_g h^4 \alt 1$;
this is to be compared with the contribution from $N_c = 3$ colors of
top quarks/squarks in the MSSM, which scales as $N_c y_t^4 \approx 3$.
Extra lepton generations can therefore help raise the Higgs mass to
its measured value only if the sleptons have extremely large masses,
leading to extra fine-tuning, which defeats one of the primary
purposes of adding a 4th generation~\cite{Martin:2009bg}. This leaves
us with only one possibility, adding a ${\bf 10}$ and any number of
{\bf 1}s.  The singlets do not impact gauge coupling unification,
cannot interact through Yukawa couplings with the Higgs boson in this
model, and do not couple to Bino dark matter, and so have no effect;
we will therefore omit them.  The resulting model, known as the QUE
model, is consistent with perturbative gauge coupling unification and
can raise the Higgs boson mass through the $H_u Q U$ interaction with
a significant Yukawa coupling.

The additional particles in the QUE model are
\begin{eqnarray}
\text{Dirac fermions:} && T_4, B_4, t_4, \tau_4 \\
\text{Complex scalars:} && \tilde{T}_{4L}, \tilde{T}_{4R}, 
\tilde{B}_{4L}, \tilde{B}_{4R}, \tilde{t}_{4L}, \tilde{t}_{4R}, 
\tilde{\tau}_{4L}, \tilde{\tau}_{4R} \ ,
\end{eqnarray}
where the subscripts 4 denote 4th generation particles, upper- and
lower-case letters denote isodoublets and isosinglets, respectively,
and $L$ and $R$ denote scalar partners of left- and right-handed
fermions, respectively.  The SUSY-preserving interactions are
specified by the superpotential
 \begin{equation}
W_{\text{QUE}} = 
M_{Q_4}\hat{Q}_4 \hat{\bar{Q}}_4 + M_{t_4} \hat{t}_4 \hat{\bar{t}}_4 
+ M_{\tau_4} \hat{\tau}_4 \hat{\bar{\tau}}_4 
+ k \hat{H}_u \hat{Q}_4 \hat{\bar{t}}_4 
- h \hat{H}_d \hat{\bar{Q}}_4 \hat{t}_4 \ ,
\label{sprptntl}
\end{equation}
where the carets denote superfields, $\hat{Q}_4 = (\hat{T}_4,
\hat{B}_4)$ is the quark isodoublet, $\hat{t}_4$ and $\hat{\tau}_4$
are the quark and lepton isosinglets, and the vector-like masses
$M_{Q_4}$, $M_{t_4}$, and $M_{\tau_4}$ and the Yukawa couplings $k$
and $h$ are all free parameters. We also assume small but
non-vanishing mixings of these fields with, say, 3rd generation
fields, so that the 4th generation fermions decay and are not
cosmologically troublesome.  These have relevance for collider
physics, but are not significant for the topics discussed here and so
are not displayed.  Finally, there are the soft SUSY-breaking terms
\begin{eqnarray}
\mathcal{L}_{\text{QUE}} &=& 
- m^2_{\tilde{Q}_4} \lvert \tilde{Q}_4 \rvert^2
- m^2_{\tilde{\bar{Q}}_4} \lvert \tilde{\bar{Q}}_4 \rvert^2
- m^2_{\tilde{t}_4} \lvert \tilde{t}_4 \rvert^2 
- m^2_{\tilde{\bar{t}}_4}\lvert \tilde{\bar{t}}_4 \rvert^2 
- m^2_{\tilde{\tau}_4}\lvert \tilde{\tau}_4 \rvert^2 
- m^2_{\tilde{\bar{\tau}}_4}\lvert \tilde{\bar{\tau}}_4 \rvert^2 
\nonumber \\ 
&& - A_{t_4} H_u \tilde{Q}_4 \tilde{\bar{t}}_4 
- A_{b_4} H_d \tilde{\bar{Q}}_4 \tilde{t}_4
- B_{Q_4}\tilde{Q}_4\tilde{\bar{Q}}_4 
- B_{t_4}\tilde{t}_4 \tilde{\bar{t}}_4 
- B_{\tau_4}\tilde{\tau}_4 \tilde{\bar{\tau}}_4 \ ,
\end{eqnarray} 
where all the coefficients are free, independent parameters.

If one drops the GUT multiplet requirement, there is another
possibility consistent with perturbative gauge coupling
unification~\cite{Martin:2009bg}: the QDEE model, with the $U$ of the
{\bf 10} replaced by a $D$, and an additional (5th generation) $E$.
This model also (accidentally) preserves gauge coupling unification
and raises the Higgs mass through the $H_d Q D$ interaction, and we
will include it in our analysis.

With notation similar to that above, the QDEE model has the extra
particles
\begin{eqnarray}
\text{Dirac fermions:} && T_4, B_4, b_4, \tau_4 , \tau_5 \\
\text{Complex scalars:} && \tilde{T}_{4L}, \tilde{T}_{4R}, 
\tilde{B}_{4L}, \tilde{B}_{4R}, \tilde{b}_{4L}, \tilde{b}_{4R}, 
\tilde{\tau}_{4L}, \tilde{\tau}_{4R} , 
\tilde{\tau}_{5L}, \tilde{\tau}_{5R} \ .
\end{eqnarray}
The superpotential is
\begin{equation}
W_{\text{QDEE}} = M_{Q_4} \hat{Q}_4 \hat{\bar{Q}}_4 
+ M_{b_4} \hat{b}_4 \hat{\bar{b}}_4 
+ M_{\tau_4} \hat{\tau}_4 \hat{\bar{\tau}}_4
+ M_{\tau_5} \hat{\tau}_5 \hat{\bar{\tau}}_5 
+ k \hat{H}_u \hat{Q}_4 \hat{\bar{b}}_4 
- h \hat{H}_d \hat{\bar{Q}}_4 \hat{b}_4 \ ,
\label{sprptntl2}
\end{equation}
and the soft SUSY-breaking terms are
\begin{eqnarray}
\mathcal{L}_{\text{QDEE}} \! \! &=& 
\! \! - m^2_{\tilde{Q}_4} \lvert \tilde{Q}_4 \rvert^2
\! \! - \! m^2_{\tilde{\bar{Q}}_4} \lvert \tilde{\bar{Q}}_4 \rvert^2
\! \! - \! m^2_{\tilde{b}_4} \lvert \tilde{b}_4 \rvert^2 
\! \! - \! m^2_{\tilde{\bar{b}}_4} \lvert \tilde{\bar{b}}_4 \rvert^2 
\! \! - \! m^2_{\tilde{\tau}_4} \lvert \tilde{\tau}_4 \rvert^2 
\! \! - \! m^2_{\tilde{\bar{\tau}}_4} \lvert \tilde{\bar{\tau}}_4 \rvert^2
\! \! - \! m^2_{\tilde{\tau}_5} \lvert \tilde{\tau}_5 \rvert^2
\! \! - \! m^2_{\tilde{\bar{\tau}}_5} 
   \lvert \tilde{\bar{\tau}}_5 \rvert^2 \nonumber \\
&& \! \! - A_{t_4} H_u \tilde{Q}_4 \tilde{\bar{b}}_4 
- A_{b_4} H_d \tilde{\bar{Q}}_4 \tilde{b}_4 
 - B_{Q_4} \tilde{Q}_4 \tilde{\bar{Q}}_4 
 - B_{b_4} \tilde{b}_4 \tilde{\bar{b}}_4 
 - B_{\tau_4} \tilde{\tau}_4 \tilde{\bar{\tau}}_4 
 - B_{\tau_5} \tilde{\tau_5} \tilde{\bar{\tau}}_5 \ .
\end{eqnarray}

\subsection{Simplifying Assumptions}
\label{sec:assumptions}

Although we have reduced the number of models we consider to two
fairly minimal ones, in each model there are still a large number of
new parameters.  To make progress and present our results, we make a
number for simplifying assumptions about the weak-scale values of
these parameters.

For both models, we choose the ratio of Higgs vacuum expectation
values to be $\tan\beta = 10$, a moderate value that makes the
tree-level Higgs mass near its maximal value.  To maximize the
radiative corrections from the 4th generation quark sector, we fix the
up-type Yukawa couplings to be at their quasi-fixed point values:
$k=1.05$ in the QUE model and 1.047 in the QDEE
model~\cite{Martin:2009bg}.  The down-type Yukawa couplings $h$ have
lower quasi-fixed point values.  They can boost the Higgs boson mass
if $h<0$, but their effects are suppressed by $\tan\beta$ and so
typically quite subdominant; for simplicity, we set $h = 0$.  We also
assume $|\mu|$ is sufficiently large that the lightest neutralino is
the Bino $\tilde{B}$.  Finally, we choose $A$-parameters such that
there is no left-right squark mixing, that is, $A_{t_4} - \mu \tan
\beta = 0$ and $A_{b_4} - \mu \cot \beta = 0$, and assume the 4th
generation $B$-parameters are negligible.

For the QUE model, we assume spectra of the extra fermions and
sfermions that can be specified by 4 parameters: the unified
(weak-scale) squark, slepton, quark, and lepton masses
\begin{eqnarray}
m_{\tilde{q}_4} &\equiv& m_{\tilde{T}_{4L}} = m_{\tilde{T}_{4R}} 
= m_{\tilde{B}_{4L}} = m_{\tilde{B}_{4R}} = m_{\tilde{t}_{4L}} 
= m_{\tilde{t}_{4R}} \\
m_{\tilde{\ell}_4} &\equiv& m_{\tilde{\tau}_{4L}} 
= m_{\tilde{\tau}_{4R}} \\
m_{q_4} &\equiv& m_{T_4} = m_{B_4} = m_{t_4} \\
m_{\ell_4} &\equiv& m_{\tau_4} \ .
\end{eqnarray}
Strictly speaking, some of these relations cannot be satisfied
exactly, as quarks (squarks) that are in the same isodoublet have
SU(2)-preserving masses specified by the same parameters, and their
physical masses are then split by electroweak symmetry breaking.
However, these splittings are small compared to the masses we will
consider and so ignoring them will have little impact on our relic
density results.

For the QDEE model, we also assume 4 unifying masses
\begin{eqnarray}
m_{\tilde{q}_4} &\equiv& m_{\tilde{T}_{4L}} = m_{\tilde{T}_{4R}} 
= m_{\tilde{B}_{4L}} = m_{\tilde{B}_{4R}} 
= m_{\tilde{b}_{4L}} = m_{\tilde{b}_{4R}} \\
m_{\tilde{\ell}_4} &\equiv& m_{\tilde{\tau}_{4L}} = m_{\tilde{\tau}_{4R}}
= m_{\tilde{\tau}_{5L}} = m_{\tilde{\tau}_{5R}} \\
m_{q_4} &\equiv& m_{T_4} = m_{B_4} = m_{b_4} \\
m_{\ell_4} &\equiv& m_{\tau_4} = m_{\tau_5} \ .
\end{eqnarray}

Finally, for both models, we assume that the Bino is lighter than all
squarks and sleptons so that it is the lightest supersymmetric
particle (LSP), but heavier than at least some fermions, so that it
can annihilate to them and reduce its thermal relic density.  For
simplicity, we assume the mass ordering
\begin{equation}
m_{\tilde{q}_4}, m_{\tilde{\ell}_4}, m_{q_4} 
> m_{\tilde{B}} > m_{\ell_4}  \ ,
\end{equation}
so that Binos annihilate to 4th generation leptons, but not 4th
generation quarks.  As we will see, the addition of the 4th generation
lepton channels is enough to reduce the Bino relic density to allowed
levels.  This ordering also allows the colored new particles to be
heavy enough to avoid LHC bounds.

\subsection{Existing Bounds}

We have included a Higgs-Yukawa term for the vector-like up-type
quarks, even though these already have vector-like masses. The
motivation, of course, is to induce corrections to the Higgs boson
mass. One has to worry, though, that such couplings could violate
electroweak constraints. In Ref.~\cite{Martin:2009bg}, however, it is
shown that already for $350~\gev$ vector-like up-type quarks, the
contributions to the $STU$ parameters are within the 1$\sigma$
exclusion contours, and the contributions are even smaller for the
heavier masses that yield the correct relic density.

Another reason one might worry about the Higgs terms is constraints
from Higgs physics, namely Higgs production and decay through triangle
diagrams with fermions in the loop. As mentioned in the introduction,
for chiral fermions, the linear relation between the fermion mass and
the Higgs Yukawa slows down the decoupling of those triangle diagrams
as the fermion mass is increased so that, by the time the experimental
constraints are satisfied, the Yukawa coupling are
non-perturbative~\cite{Ellis:2014dza}. Adding a vector-like mass makes
these triangle diagrams decouple more quickly.  However, there are
still some limits from the LHC Higgs data, which we take from
Ref.~\cite{Ellis:2014dza}. According to their analysis, vector-like
quarks of about 1 TeV are (barely) safe from experimental limits. Note
however that their fit is based on a model with both up- and down-type
isosinglets, so their limits will be weaker when applied to our
models, where either the down-type or up-type isosinglet is
missing. The authors also perform a fit to the $STU$ parameters that
confirms our conclusions based on Ref.~\cite{Martin:2009bg} that our
model is safe.

Last, as noted above, to allow the 4th generation fermions to decay
and so satisfy cosmological bounds, we assume that they mix with MSSM
fields.  In general, the 4th generation fields may then induce
magnetic or electric dipole moments or mediate flavor-violating
observables for fermions in the first 3 generations.  We will assume
that these mixings are minute, however, and dominantly with the 3rd
generation, where bounds are weak and easily consistent with the
lifetime requirement from cosmology.

\section{Relic Density}
\label{sec:relic}

With the assumptions of \secref{assumptions}, there are now new dark
matter annihilation processes: $\tilde{B} \tilde{B} \to \tau_i^+
\tau_i^-$, mediated by $t$- and $u$-channel sleptons
$\tilde{\tau}_{iL}$ and $\tilde{\tau}_{iR}$, where $i=4$ for the QUE
model and $i=4,5$ for the QDEE model.  These new channels increase the
thermally-averaged annihilation cross section $\langle \sigma v
\rangle$, which may reduce the Bino thermal relic density
$\Omega_{\tilde{B}} h^2$ to acceptable levels even for large and
viable Bino masses.

For the present purposes, it suffices to calculate the relic density
using the approximation~\cite{kolbturner}
\begin{eqnarray}
\Omega_{\tilde{B}} h^2 &=& 1.07\times 10^{9}~\gev^{-1}
\frac{x_{f}}{\sqrt{g_*} \, \mplanck \, a \, [ 1 + b / (2a x_F) ] } 
\label{omega}  \\
x_F &=& \ln r -  \frac{1}{2} \ln \left(\ln r \right) 
+ \ln \left(1 + b / \ln r \right) \label{xf} \\ 
r &=& 0.038 \frac{g}{\sqrt{g_{*} \mplanck m_{\chi} a}} \ ,
\end{eqnarray}
where $x_F = m_{\tilde{B}}/T_F$, the ratio of the dark matter mass to
the freezeout temperature $T_F$, $g_*$ is the number of massless
degrees of freedom at freezeout, $g=2$ is the number of degrees of
freedom of the Bino, $\mplanck \simeq 1.22 \times 10^{19}~\gev$ is the
Planck mass, and $a$ and $b$ are the $S$- and $P$-wave cross section
coefficients given below.  For the parameters of interest here, we
find $x_F \approx 24$, and so $T_F$ is between the $W$ and $b$ masses
and $g_* \approx 87.25$.  The current bound on the dark matter relic
density is $\Omega_{\text{DM}} h^2 = 0.1199 \pm
0.0022$~\cite{Ade:2015xua}. \Eqref{omega} is accurate to
5\%~\cite{kolbturner} or better, and we will require
$\Omega_{\tilde{B}} h^2 = 0.12$ to within a fractional accuracy of
10\%.

The cross section for $\tilde{B} \tilde{B} \to f^+f^-$ mediated by
$t$- and $u$-channel sfermions $\tilde{f}_{L,R}$ with masses $m_{L,R}$
and hypercharges $Y_{L,R}$ is
\begin{eqnarray}
\left. \frac{d\sigma}{d\Omega} \right|_{\text{CM}}
&=& \frac{1}{256 \pi^2 s} \sqrt{\frac{s-4 m_f^2}{s - 4 m_{\tilde B}^2}}
\sum_{i,f} \left| \cal{M} \right|^2 \\
\sum_{i,f} \left| \cal{M} \right|^2 
&=& \frac{1}{4} g_Y^4 Y_L^4 \left[
\frac{( m_{\tilde B}^2 + m_f^2 - t)^2}{(m_L^2 - t)^2}
+ \frac{( m_{\tilde B}^2 + m_f^2 - u)^2}{(m_L^2 - u)^2}
- \frac{ 2 m_{\tilde B}^2 (s - 2 m_f^2) } { (m_L^2 - t) (m_L^2 - u)} 
\right] \nonumber \\
&& +  \frac{1}{4} g_Y^4 Y_R^4 \left[
\frac{( m_{\tilde B}^2 + m_f^2 - t)^2}{(m_R^2 - t)^2}
+ \frac{( m_{\tilde B}^2 + m_f^2 - u)^2}{(m_R^2 - u)^2}
- \frac{ 2 m_{\tilde B}^2 (s - 2 m_f^2) } { (m_R^2 - t) (m_R^2 - u)} 
\right] \nonumber \\
&& + \frac{1}{2} g_Y^4 Y_L^2 Y_R^2 m_f^2 \left[
\frac{4 m_{\tilde B}^2}{(m_L^2 - t) (m_R^2 - t)}
+ \frac{4 m_{\tilde B}^2}{(m_L^2 - u) (m_R^2 - u)} \right. \nonumber \\
&& \qquad \qquad \qquad \quad
\left. - \frac{s - 2 m_{\tilde B}^2}{(m_L^2 - t) (m_R^2 - u)}
- \frac{s - 2 m_{\tilde B}^2}{(m_L^2 - u) (m_R^2 - t)} \right] ,
\end{eqnarray}
where $g_Y \simeq 0.35$ is the U(1)$_Y$ gauge coupling. 

Multiplying this differential cross section by the relative velocity
$v$, expanding in powers of $v$, integrating over angles, and carrying
out the thermal average, we find
\begin{eqnarray}
\langle \sigma v \rangle &=& a + b\, x_F^{-1} 
\label{xsection} \\
a &=& \frac{g^{4}_{Y}}{128 \pi }\frac{m_{f}^{2}}{m_{\tilde{B}}}  
\sqrt{m_{\tilde{B}}^{2}-m_{f}^{2}} 
\left [\frac{Y_{L}^{4}}{\Delta_{L}^{2}}    
+ \frac{Y_{R}^{4}}{\Delta_{R}^{2}} 
+ \frac{2Y_{L}^{2}Y_{R}^{2}}{\Delta_{L}\Delta_{R}} \right ] 
\label{xsection1} \\
b &=& \frac{g^{4}_{Y}}{512 \pi }
\frac{1}{m_{\tilde{B}}\sqrt{m_{\tilde{B}}^{2}-m_{f}^{2}}} 
\left] \frac{Y_{L}^{4}}{\Delta_{L}^{4}} f_{LL}
+ \frac{Y_{R}^{4}}{\Delta_{R}^{4}} f_{RR} 
+ \frac{Y_L^2 Y_R^2}{\Delta_L \Delta_R} m_f^2 f_{LR} \right] \ ,
\end{eqnarray}
where
\begin{eqnarray}
f_{LL,RR} &=& 
13 m_f^8 + m_f^6 \left(-26 m_{L,R}^2 - 36 m_{\tilde{B}}^2\right) 
+m_f^4 \left(70 m_{L,R}^2 m_{\tilde{B}}^2 +13 m_{L,R}^4
 + 49 m_{\tilde{B}}^4\right) \nonumber   \\
&& +  m_f^2 \left(-44 m_{L,R}^2 m_{\tilde{B}}^4 
- 26 m_{L,R}^4 m_{\tilde{B}}^2 - 42 m_{\tilde{B}}^6 \right) 
+ 16 \left(m_{L,R}^4 m_{\tilde{B}}^4 + m_{\tilde{B}}^8\right) \\
f_{LR} &=&
\left( 18 m_{f}^{2}-12 m_{\tilde{B}}^{2}\right ) \nonumber \\
&& + \frac{8  \left(m_{\tilde{B}}^2-m_f^2\right) }
{\Delta_{L}^{2}\Delta_{R}^{2}}  
\left[ -3 m_f^8+m_f^6 \left( 8 m_{\tilde{B}}^2 
+ 6 m_L^2 + 6 m_R^{2}\right) \right. \nonumber \\
&& +   m_f^4 \left(-6 m_{\tilde{B}}^4-17 m_L^2 m_{\tilde{B}}^2 
- 3 m_L^4-17 m_R^2 m_{\tilde{B}}^2-3 m_R^4-12 m_L^2 m_R^2 \right) \nonumber \\
&& +   m_f^2 \left(6 m_L^4 m_R^2+7 m_L^4 m_{\tilde{B}}^2
+ 16 m_L^2 m_{\tilde{B}}^4 \right. \nonumber \\
&& \left. +6 m_R^4 m_L^2+7 m_R^4 m_{\tilde{B}}^2
+ 16 m_R^2 m_{\tilde{B}}^4
+ 30 m_L^2 m_R^2 m_{\tilde{B}}^2\right) \nonumber \\
&&+ m_{\tilde{B}}^8-5 m_L^2 m_{\tilde{B}}^6-4 m_L^4 m_{\tilde{B}}^4 
- 9 m_L^2 m_R^4 m_{\tilde{B}}^2 \nonumber \\
&&\left. -5 m_R^2 m_{\tilde{B}}^6-4 m_R^4 m_{\tilde{B}}^4 
- 9 m_R^2 m_L^4 m_{\tilde{B}}^2 - 3 m_L^4 m_R^4 
- 18 m_L^2 m_R^2 m_{\tilde{B}}^4 \right] \ ,
\label{xsection2}
\end{eqnarray}
and $\Delta_{L,R}= m^{2}_{\tilde{B}}+m_{L,R}^{2}-m_{f}^{2}$. 

Equations~(\ref{xsection})--(\ref{xsection2}) are valid for sfermions
with different masses and hypercharges.  For degenerate vector-like
sfermions with $m_{\tilde{f}} \equiv m_L = m_R$ and $Y_V \equiv
Y_L=Y_R$, the cross section coefficients simplify to
\begin{eqnarray}
a &=& \frac{g^{4}_{Y} Y_{V}^{4}}{32 \pi}\frac{m_{f}^{2}}{m_{\tilde{B}}}
\frac{\sqrt{m_{\tilde{B}}^{2}-m_{f}^{2}}}
{ \left (m_{\tilde{B}}^{2}+m_{\tilde{f}}^{2}-m_{f}^{2}\right )^{2}} 
\label{xsectionsimple1} \\
b &=& \frac{g^{4}_{Y} Y_{V}^{4}}{128 \pi }\frac{1}{m_{\tilde{B}}} 
\frac{1}{\sqrt{m_{\tilde{B}}^{2}-m_{f}^{2}} 
\left (m_{\tilde{B}}^{2}+m_{\tilde{f}}^{2}-m_{f}^{2}\right )^{4}} \times
\nonumber \\
&& \left[17 m_f^8-2 m_f^6 
\left(17 m_{\tilde{f}}^2+20 m_{\tilde{B}}^2\right)
+ m_f^4 \left( 86 m_{\tilde{B}}^2 m_{\tilde{f}}^2 + 17 m_{\tilde{f}}^4
+ 37 m_{\tilde{B}}^4\right) \right. \nonumber \\
&& \quad - \left. 2 m_f^2 \left(26 m_{\tilde{B}}^4 m_{\tilde{f}}^2 
+ 11 m_{\tilde{B}}^2 m_{\tilde{f}}^4+11 m_{\tilde{B}}^6\right)
+ 8 m_{\tilde{B}}^4 \left(m_{\tilde{f}}^4+m_{\tilde{B}}^4\right) \right] \ .
\label{xsectionsimple2}
\end{eqnarray}

The expansion in $v$ assumes that $v$ is the only small parameter.
This is not true when $f$ and $\tilde{B}$ become degenerate and the
annihilation is near threshold.  In this limit, the expressions for
$b$ in \eqsref{xsection2}{xsectionsimple2} become singular, signaling
the breakdown of the expansion.  The expansion is essentially an
expansion in even powers of $\alpha = v / \sqrt{1 - (m_f
  /m_{\tilde{B}})^2}$.  Requiring that the next omitted ($D$-wave)
term be less than a 10\% correction implies roughly $\alpha^4 < 0.1$.
For characteristic velocities of $v \sim 0.3$ at freezeout, this
implies $m_f < 0.85 m_{\tilde{B}}$.  The case of near-threshold
annihilation was considered in Ref.~\cite{Griest:1989zh}, where it was
shown in a generic setting that corrections to $\langle \sigma v
\rangle$ above the few percent level may occur if $m_f > 0.95
m_{\tilde{B}}$. There, alternative expressions valid in the degenerate
limit were derived.  Here, as we are primarily interested in the
cosmologically preferred regions without accidental mass degeneracies,
we will use \eqsref{xsectionsimple1}{xsectionsimple2} and simply take
care to avoid applying these cross section formulae to cases where the
dark matter and final state fermion are in the degenerate region.  We
note also that an expression for $\langle \sigma v \rangle$ was
presented in Ref.~\cite{Olive:1989jg} for degenerate sfermions. The
expressions there differ from our result in \eqref{xsection} with $m_L
= m_R$, but the disagreement is numerically small and at most at the
5\% level.

The annihilation cross section has some interesting features.  First,
hypercharge enters to the fourth power.  Isosinglet leptons have the
largest hypercharge of any MSSM fields.  As we show below, the squarks
need to be above a TeV to achieve the correct Higgs mass.  But leptons
and sleptons can be relatively light.  As a result, annihilation to
leptons is particularly efficient, and it is fortunate that they exist
in both the QUE and QDEE models. Note also that because the fermions
are vector-like, there is no chiral suppression.  This differs greatly
from the MSSM, where annihilations to isosinglet leptons are
hypercharge-enhanced, but extremely suppressed by the chiral
suppression of the $S$-wave cross section, since all MSSM leptons are
light.  In both the QUE and QDEE models, there are heavy isosinglet
leptons, and annihilation to them is neither hypercharge- nor
chirality-suppressed.  Annihilations to 4th generation particles
therefore completely dominate over MSSM channels.

\begin{figure}[t]
\includegraphics[width=.48\linewidth]{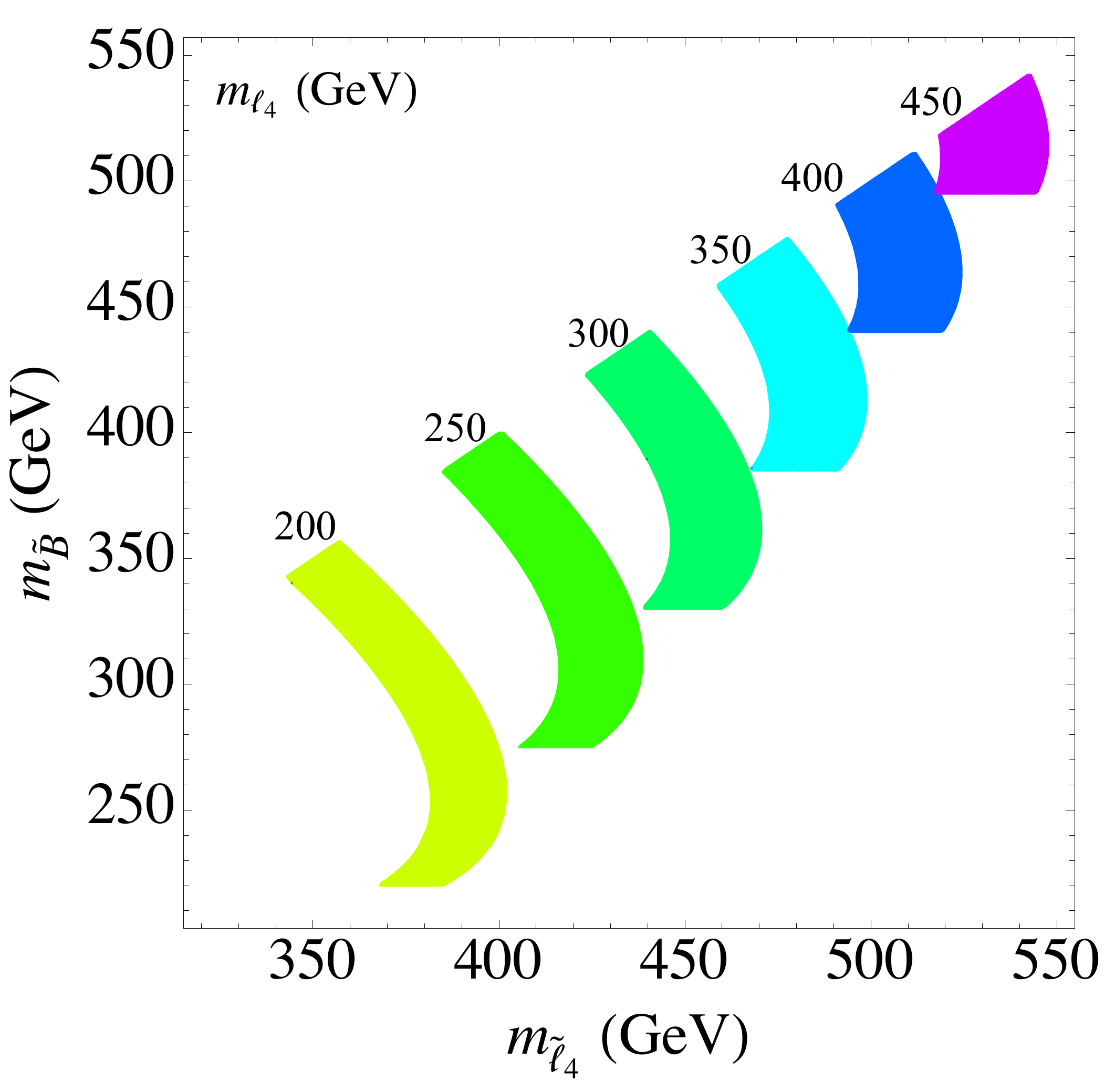} \
\includegraphics[width=.48\linewidth]{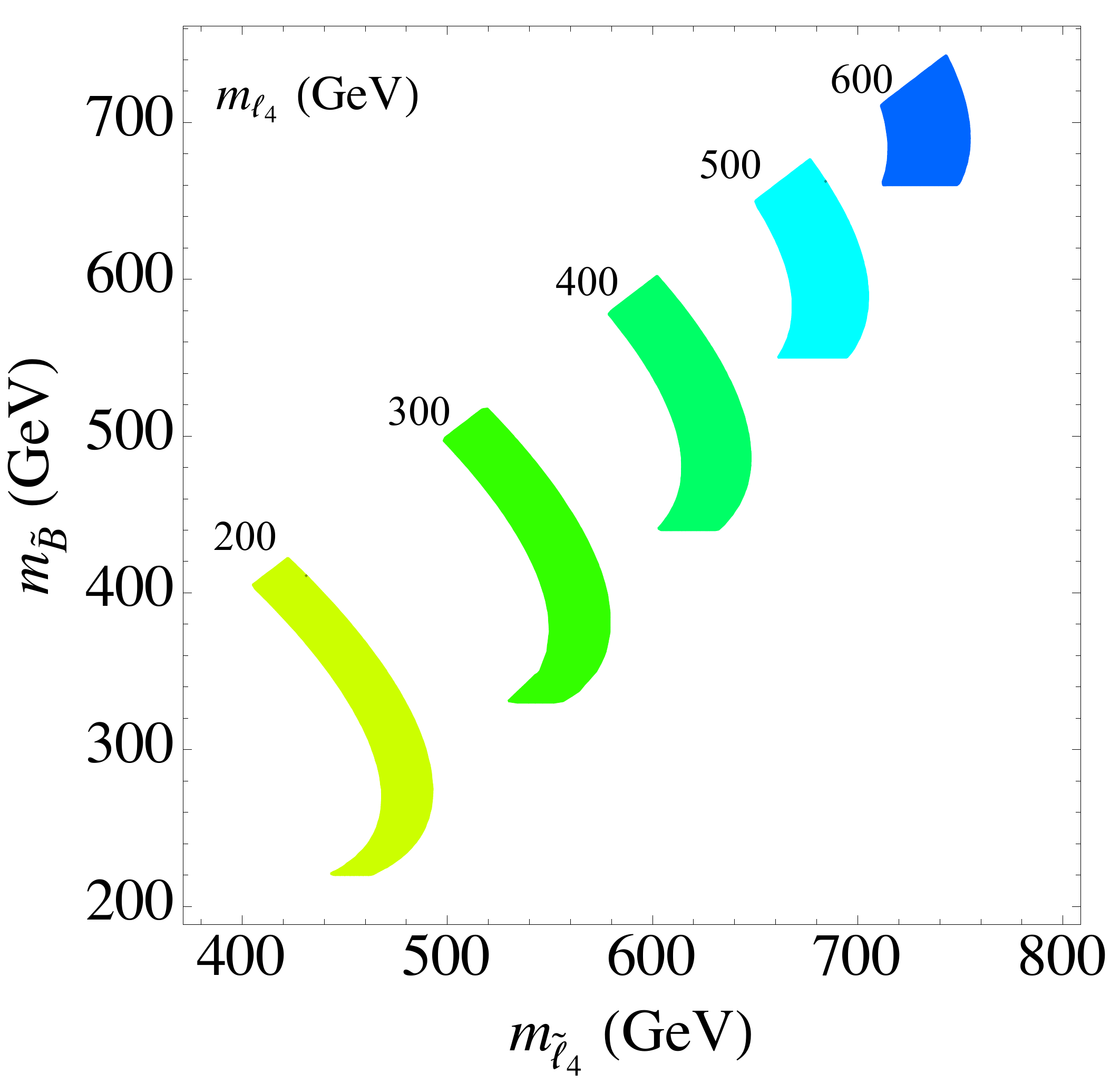}
\vspace*{-0.1in}
\caption{Cosmologically preferred regions in the $(m_{\tilde{\ell}_4},
  m_{\tilde{B}})$ plane for the QUE (left) and QDEE (right) models.
  In each shaded region, the relic density is in the preferred range
  $\Omega_{\tilde{B}} h^2 = 0.12 \pm 0.012$ for the value of
  $m_{\ell_4}$ indicated. 
\label{fig:Rlc1} }
\vspace*{-0.1in}
\end{figure}

In \figref{Rlc1}, we show regions of the $(m_{\tilde{\ell}_4},
m_{\tilde{B}})$ plane, with $m_{\ell_4}$ fixed to the values
indicated, where Bino dark matter freezes out with a relic density
within 10\% of the value required to be all of dark matter.  These
regions are bounded on all sides.  We must require the mass ordering
$m_{\ell_4} < m_{\tilde{B}} < m_{\tilde{\ell}_4}$ so that the Binos
are the LSPs, but may pair-annihilate to 4th generation leptons.  The
mass of $\ell_4$ is bounded from below by heavy lepton searches. As
this mass is increased, the Bino and slepton masses must also increase
to maintain the mass ordering.  As the masses increase, however, the
annihilation cross section $\langle \sigma v \rangle$ decreases, and
at some point the thermal relic density of Binos is too large,
providing an upper bound on all of these masses.  To guarantee that the 
velocity expansion is reliable in the regions shown in \figref{Rlc1}, we 
have required $m_{\ell_4} < 1.1m_{\tilde{B}} $.  We have not 
included co-annihilation, which would be important for 
Binos and sleptons that are degenerate to more than 5\%. 

\begin{figure}[t]
\includegraphics[width=.48\linewidth]{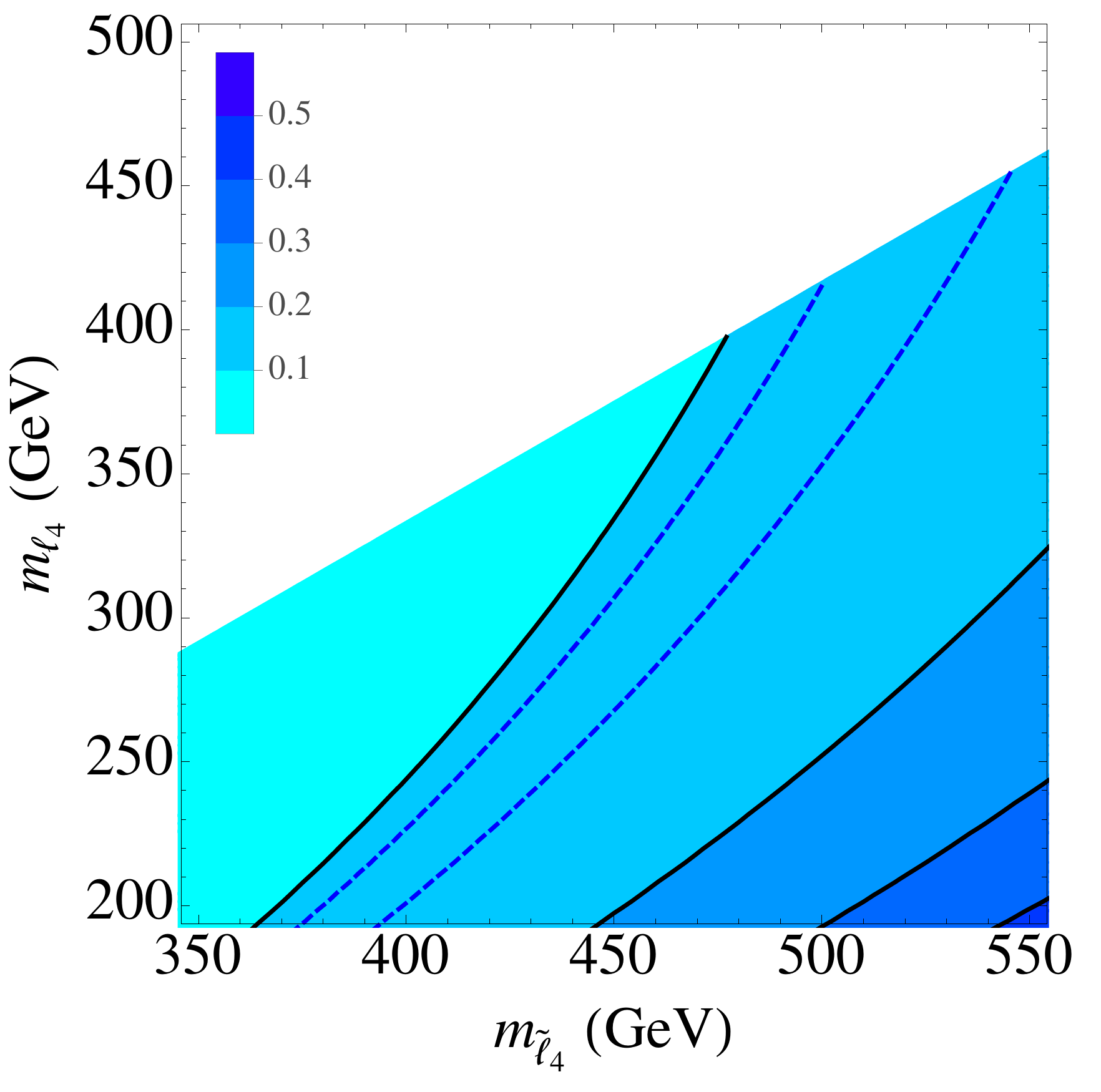} \
\includegraphics[width=.48\linewidth]{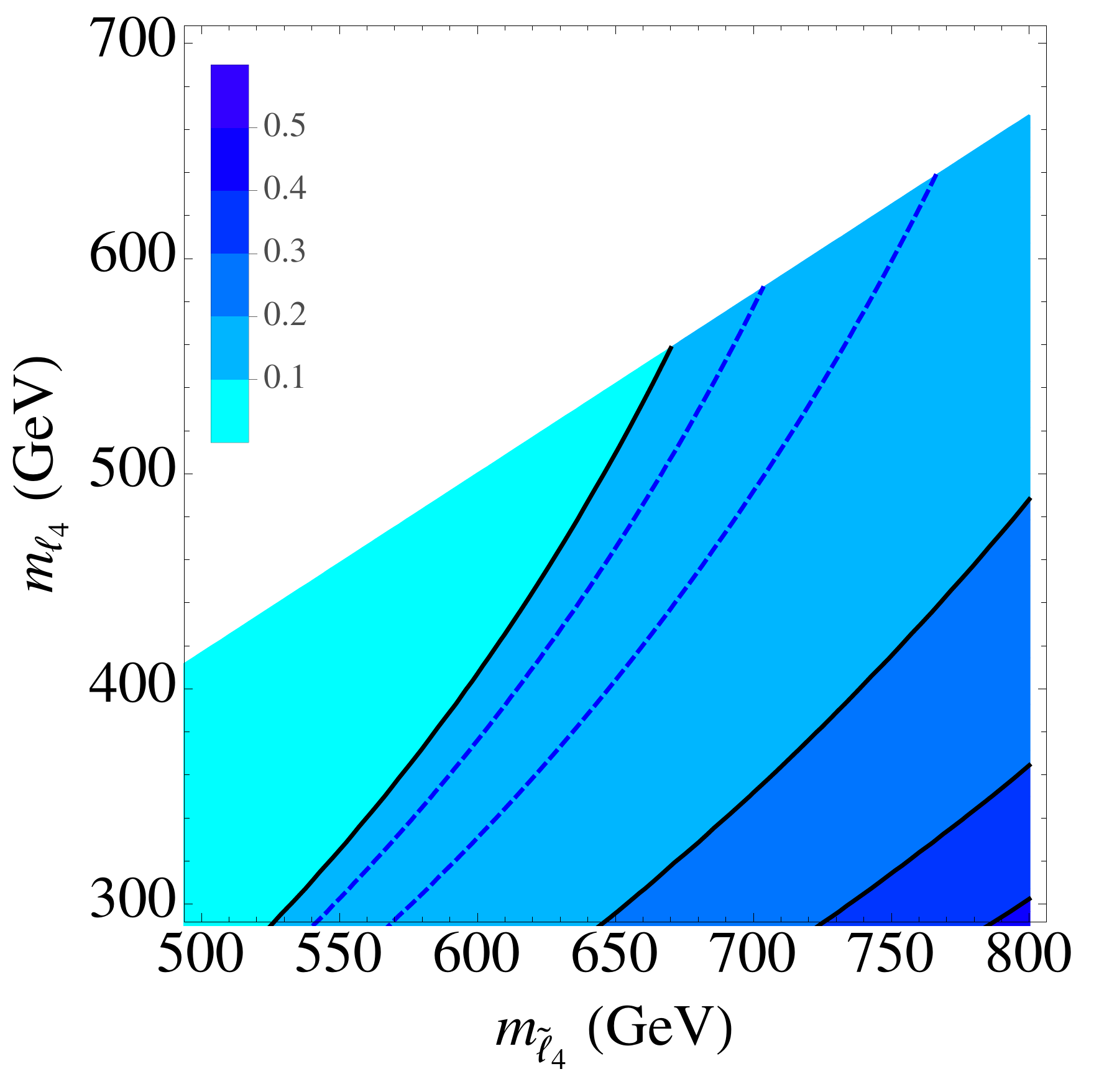}
\vspace*{-0.1in}
\caption{Contours of constant relic density $\Omega_{\tilde{B}} h^2$
  for the QUE (left) and QDEE (right) models in the $(m_{{\tilde
      \ell}_4}, m_{\ell_4})$ plane with fixed $m_{\tilde{B}} = 1.2
  m_{\ell_4}$. Between the dashed lines $\Omega_{\tilde{B}} h^2 = 0.12
  \pm 0.012$.
  \label{fig:Rlc2} }
\vspace*{-0.1in}
\end{figure}

Without co-annihilation, the largest possible masses are about
$m_{\ell_4} = 470~\gev$ in the QUE model and $670~\gev$ in the QDEE
model. To see this upper bound more clearly, in \figref{Rlc2} we plot
the relic density bands in the $(m_{{\tilde \ell}_4}, m_{\ell_4})$
plane for fixed $m_{\tilde{B}} = 1.2 m_{\ell_4}$.  Larger masses are
allowed in the QDEE model, because there are two new annihilation
channels, and since $\langle \sigma v \rangle \sim m^{-2}$, the upper
bound on the masses is larger by roughly a factor of $\sqrt{2}$.

In \figsref{Rlc1}{Rlc2} we have completely neglected the MSSM
annihilation channels; including them would only move the preferred
regions to slightly lower masses.  For the reasons mentioned above,
vector-like 4th generation particles are extremely efficient channels
for annihilation and completely dominate the MSSM contributions in the
case of Bino dark matter.  As a result, the cosmologically-preferred
Bino masses are significantly higher than in the MSSM and completely
eliminate the tension between the relic density constraints and
current LHC bounds on neutralino masses.

\section{Higgs Boson Mass}
\label{sec:higgs}

In the MSSM, the Higgs boson mass is maximally the $Z$ boson mass $M_Z
= 91~\gev$ at tree level, but is raised by radiative corrections,
dominantly from the diagrams with top quarks and squarks in loop.  Up
to 2-loop corrections, assuming no left-right stop mixing, the Higgs
boson mass is~\cite{Carena:1995bx}
\begin{equation}
\label{hgs2lp}
m_h^2 = M_Z^2 \cos^2 2\beta 
\left(1 - \frac{3}{8\pi^2} \frac{m_t^2}{v^2} t \right)
+ \frac{3}{4\pi^2} \frac{m_t^4}{v^2} 
\left[ t+\frac{1}{16\pi^2} \left( \frac{3}{2} \frac{m_t^2}{v^2} 
-32 \pi \alpha_3 \right) t^2 \right]  \ , 
\end{equation}
where 
\begin{eqnarray}
t &=& \ln \frac{M^2_{\tilde{t}}}{M_t^2} \\
m_t &=&\frac{M_{t}}{1+\frac{4}{3\pi}\alpha_{3}(M_{t})} \\
\alpha_3 &=& \frac{\alpha_{3}(M_Z)}{1 + \frac{b_3}{4\pi}
\alpha_3 (M_Z) \ln (M_t^2 / M_Z^2)} \\
b_3 &=& 11 - 2 N_f / 3 = 7  \ ,
\end{eqnarray}
$M_t = 174~\gev$ is the top quark mass, $M_{\tilde{t}}$ characterizes 
the masses of the left- and right-handed top squarks, $v=174~\gev$ 
is the Higgs vacuum
expectation value, $\alpha_{3}(M_Z) = 0.118$ is the strong gauge
coupling at the $Z$ pole, and $b_3$ is the beta coefficient for the
strong coupling in the MSSM without the top quark and any extra
matter.  For $\tan \beta = 10$, the tree-level mass is near its
maximal value, but even with top squark masses $M_{\tilde{t}} =
2~\tev$, the Higgs mass is only 115 GeV, far short of the measured
value of 125 GeV.

With the addition of vector-like quarks, however, this mass can be
significantly increased. The contribution from a vector-like 4th
generation of top quarks and squarks
is~\cite{Babu:2008ge,Martin:2009bg}
\begin{equation}
\label{hgs4g}
\Delta m_{h}^2 = \frac{N_{c} v^2}{4\pi^2} 
\left(k \sin\beta\right)^4f(x)  \ ,
\end{equation}
 where $N_{c}=3$ is the number of colors, $k$ is the up-type Yukawa
coupling in \eqsref{sprptntl}{sprptntl2}, and
\begin{eqnarray}
f(x)&=& \ln x - \frac{1}{6} \left( 5-\frac{1}{x} \right)
\left(1-\frac{1}{x}\right) \\
x&=& \frac{m_{\tilde{q}_4}}{m_{q_4}} \ .
\end{eqnarray}
As a reminder, $m_{q_4}$ and $m_{\tilde{q}_4}$ are the physical masses
of the 4th generation quarks and squarks, respectively, and we set $k$
at its quasi-fixed point value $k=1.05$ and neglect the 4th generation
down-type Yukawa $h$. Note that we are also neglecting 2-loop
contributions from vector-like matter, since those contributions are
small for $m_{q_4}, m_{\tilde{q}_4} \gg m_h$~\cite{Martin:2009bg}.

\begin{figure}[t]
\includegraphics[width=.60\linewidth]{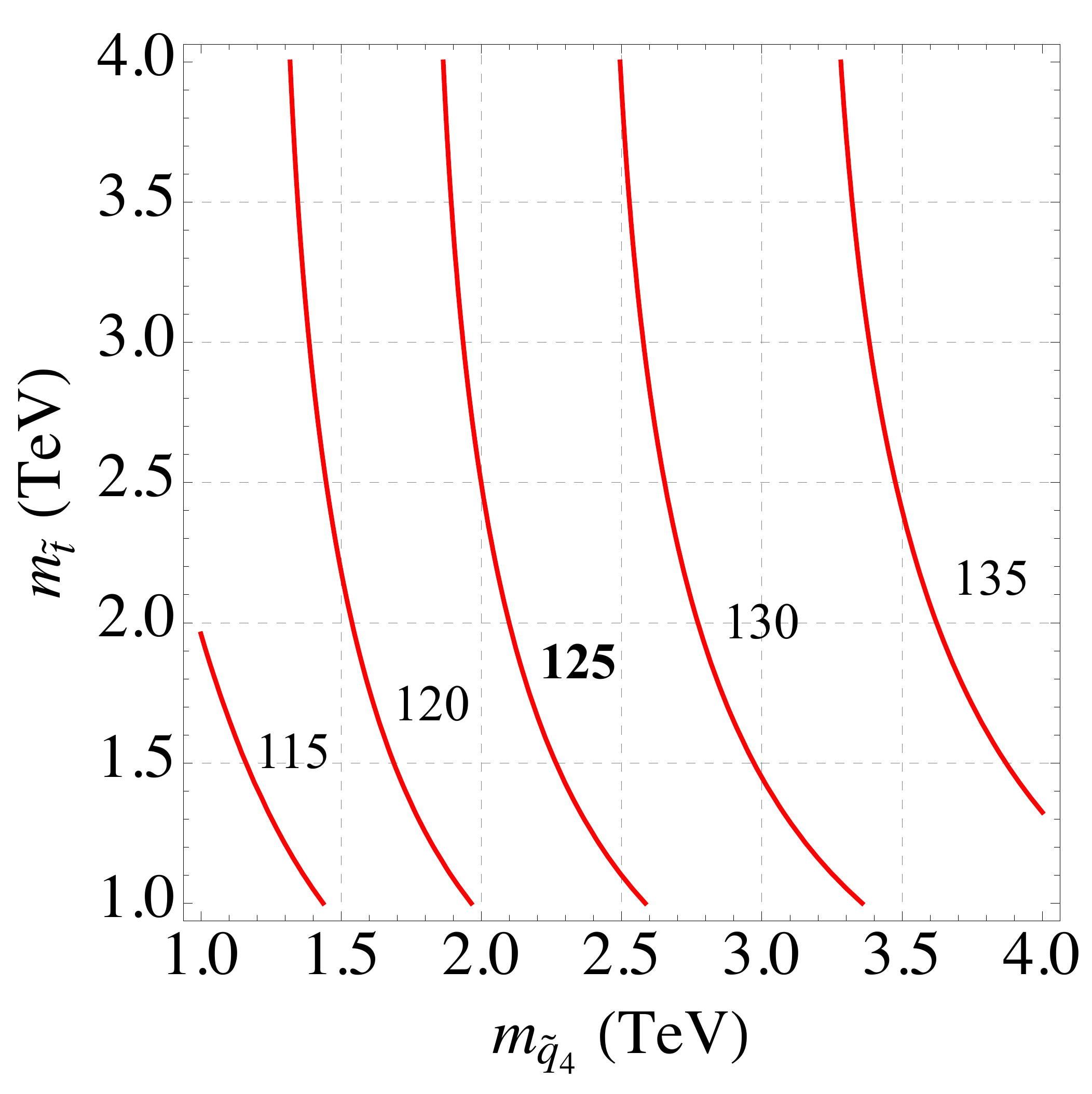}
\vspace*{-0.1in}
  \caption{Contours of constant Higgs boson mass (in GeV) in the 
  $(m_{\tilde{q}_4}, m_{\tilde{t}})$ plane, assuming no left-right
  squark mixings, for fixed $m_{t_4} = 1~\tev$.
  \label{fig:HgsCntrs} } 
\vspace*{-0.1in}
\end{figure}

We can see from \eqref{hgs4g} that the 4th generation contribution to
the Higgs boson mass is maximal when $q_4$ is as light as possible.
In \figref{HgsCntrs} we show contours of the Higgs mass in the
$(m_{\tilde{t}_4}, m_{\tilde{t}})$ plane for fixed $m_{t_4} =
1~\tev$. This choice of $m_{t_4}$ is based partly on the $\sim
700~\gev$ limit on chiral 4th generation up-type
quarks~\cite{Agashe:2014kda} and partly on the $STU$ and Higgs
constraints mentioned earlier. We see that, with the addition of 4th
generation tops, the correct Higgs mass can be achieved for a range of
$m_{\tilde{t}_4}$ and $m_{\tilde{t}}$ where both are below 3 TeV and
discoverable at future runs of the LHC.  One can see from
\eqref{hgs4g} that the corrections from the vector-like matter are
functions of $x = m_{\tilde{t}_4}/m_{t_4}$. One can use this to
reinterpret \figref{HgsCntrs} as determining the required ratio $x$ to
get the correct mass. For example, for $m_{t_4}$ between 1 and 2 TeV,
the correct Higgs boson mass can be obtained as long as $x$ is between
2.5 and 2.

\section{Conclusions}
\label{sec:end}

In this work, we have considered the cosmology of MSSM4G models, in
which the MSSM is extended by adding vector-like 4th (and 5th)
generation particles.  Remarkably, requiring perturbative gauge
coupling unification and that the extra particles raise the Higgs
boson mass significantly reduces the number of MSSM4G models to two:
the QUE and QDEE models.

Here we have shown that these models accommodate an excellent dark
matter candidate, the Bino.  In the MSSM, Bino dark matter must be
lighter than 300 GeV to avoid overclosing the Universe.  Such light
Binos are in tension with constraints from the LHC in many scenarios.
In contrast, in the MSSM4G models, Binos may annihilate to extra
leptons through $\tilde{B} \tilde{B} \to \ell_i^+ \ell_i^-$, where
$i=4$ in the QUE model, and $i=4,5$ in the QDEE model.  These
annihilation channels are enhanced by the large hypercharges of lepton
isosinglets, are not chirality-suppressed, and completely dominate
over all of the MSSM annihilation channels combined.  We have shown
that these extra channels enhance the annihilation cross section to
allow Bino masses as large as 470~GeV and 670~GeV in the QUE and QDEE
models, respectively, without requiring co-annihilations or
resonances.  MSSM4G models are therefore motivated by dark matter
also, as they accommodate Bino dark matter with the correct relic
density in completely generic regions of parameter space.

An interesting question is how to discover supersymmetry if these
MSSM4G models are realized in nature.  As we have discussed, these
models satisfy precision constraints from Higgs boson properties,
electroweak physics, and low-energy observables; future improvements
in these areas could see hints of anomalies from 4th generation
particles, but this is not generic.  These models also have improved
naturalness relative to the MSSM, in the sense that the top squarks
and 4th generation quarks and squarks, even without left-right mixing,
may be lighter than 2 to 3 TeV and still give the correct Higgs boson
mass.  These are within reach of future runs of the LHC. As noted in
\secref{higgs}, however, it is also possible for the stop and 4th
generation quarks and squarks to all be beyond the reach of the LHC.

However, the relic density does imply upper bounds on the masses of
the 4th generation leptons and sleptons. Given this, it is very
interesting to see how one could best search for these at both hadron
and lepton colliders.  Of course, Bino dark matter can also be
searched for through direct and indirect dark matter searches.  We
plan to evaluate the efficacy of these searches in a future
study~\cite{workinprogress}.

Last, we note that there are many variations one could consider.  We
have assumed many mass unifications to simplify the presentation of
our results; these could be relaxed. One could also contemplate
left-right mixings for the squarks and their impact on the Higgs boson
mass, or allow the lightest neutralino to include Higgsino or Wino
components.  We believe that the essential point is clear, though: the
combination of supersymmetry and vector-like fourth generation
particles accommodates an excellent Bino dark matter candidate even in
its simplest realizations, and the QUE and QDEE models are among the
more motivated and viable supersymmetric extensions of the standard
model.

\section*{Acknowledgments}
 
We thank Michael Geller, Danny Marfatia, and Sho Iwamoto for useful
conversations.  This work is supported in part by NSF Grant
No.~PHY--1316792. The work of J.L.F.\ was performed in part at the
Aspen Center for Physics, which is supported by NSF Grant
No.~PHY--1066293.  J.L.F. was supported in part by a Guggenheim
Foundation grant and in part by Simons Investigator Award \#376204.

\providecommand{\href}[2]{#2}\begingroup\raggedright
\endgroup


\end{document}